%% file: Zcs-phys-rev6.tex
\documentclass[aps,prd,twocolumn,groupedaddress,nofootinbib]{revtex4-1}
\usepackage{graphicx}
\usepackage{here}
\usepackage[dvips]{epsfig}
\def\beq{\begin{equation}}
\def\eeq{\end{equation}}
\def\bea{\begin{eqnarray}}
\def\eea{\end{eqnarray}}
\def\bq{\begin{quote}}
\def\eq{\end{quote}}

\def\nnb{\nonumber}
\def\ga{\left(}
\def\dr{\right)}

\def\nnb{\nonumber}

\def\la{\langle}
\def\ra{\rangle}

\def\ba{\vspace*{-0.2cm}\begin{array}}
\def\ea{\end{array}\vspace*{-0.2cm}}

\def\als{\alpha_s}

\def\gg2{\la\alpha_s G^2 \ra}
\def\gg3{g^3f_{abc}\la G^aG^bG^c \ra}
\def\ggg4{\la\als^2G^4\ra}

\def\gg{\lag g^{2}_{s} G^2 \rag}
\def\ggg{\lag g^{3}_{s}G^3\rag}

\usepackage{amsmath}
\usepackage{slashed}
\usepackage{color}

\begin{document}

\title{Tests of the $Z_c$-like Laplace Sum Rule (LSR) results using FESR at NLO
}
\author{R.M. Albuquerque}
\affiliation{Faculty of Technology, Rio de Janeiro State University (FAT,UERJ), Brazil}
\email[Email address:~] {raphael.albuquerque@uerj.br}
\author{S. Narison
}
\altaffiliation{ICTP-Trieste consultant for Madagascar - Corresponding author.}
\affiliation{Laboratoire Univers et Particules de Montpellier (LUPM) \\
CNRS-IN2P3 and University of Montpellier  
Case 070, Place Eug\`ene
Bataillon, 34095 - Montpellier, France\\
and\\
Institute of High-Energy Physics of Madagascar (iHEPMAD)\\
University of Ankatso,
Antananarivo 101, Madagascar}
\email[Email address:~] {snarison@yahoo.fr} 

\author{D. Rabetiarivony}
\email[Email address:~] {rd.bidds@gmail.com}

\affiliation{Institute of High-Energy Physics of Madagascar (iHEPMAD)\\
University of Ankatso,
Antananarivo 101, Madagascar}

\date{\today}
\begin{abstract}
\noindent
In this note, we use local duality Finite Energy Sum Rule (FESR) to test the
validity of the Laplace sum rules (LSR) results truncated at the dimension-six condensates for the estimates of the masses and couplings of 
the $Z_c$-like ground states in Ref.\,\cite{Zc} by taking the example of the $D^*D$ molecule configuration. We confirm the existence of an eventual $(D^*D)_1$ radial excitation with a mass  around 5700 MeV and coupling of 197(25) keV to the current which may mask the eventual $Z_c(4430)$ radial excitation candidate (named $(D^*D)_0$ in Ref.\,\cite{Zc}) having a relatively small coupling $f_{(D^*D)_0}$=46(56) keV.  We add more explanations on the estimates in Ref.\,\cite{Zc}
from LSR and comment the results in Ref\,\cite{WANG}. 
\end{abstract}
\maketitle
 \section{Introduction}
 In Ref.\,\cite{Zc}, we have estimated 
 the masses and couplings of $Z_c$-like states within different configurations  of their eventual nature  using Laplace sum rules 
 (LSR)\,\cite{BELLa,BERTa,BECCHI,SNR} \`a la SVZ\,\cite{SVZa,ZAKA} and their ratios at NLO of PT series:
\bea
 {\cal L}^c_n(\tau,\mu)&=&\int_{t_0}^{t_c}\hspace*{-0cm}dt~t^n~e^{-t\tau}\frac{1}{\pi} \mbox{Im}~\Pi^{(1)}_{\cal H}(t,\mu)~,\nnb\\
 {\cal R}^c_n(\tau)&=&\frac{{\cal L}^c_{n+1}} {{\cal L}^c_n}.
\label{eq:lsr}
\eea
$m_c$ is the charm quark mass, $\tau$ is the LSR variable, $n=0,1$ is the degree of moments, $t_0$ is the quark/hadronic threshold. $t_c$ is the threshold of the ``QCD continuum" which parametrizes, from the discontinuity of the Feynman diagrams, the spectral function  ${\rm Im}\,\Pi^{(1)}_{\cal H}(t,m_c^2,\mu^2)$.   $\Pi^{(1)}_{\cal H}(t,m_c^2,\mu^2)$ is the  transverse scalar correlator corresponding to a spin one hadron\,: 
 \bea
\hspace*{-0.6cm} \Pi^{\mu\nu}_{\cal H}(q^2)&=&i\int \hspace*{-0.15cm}d^4x ~e^{-iqx}\la 0\vert {\cal T} {\cal O}^\mu_{\cal H}(x)\ga {\cal O}^\nu_{\cal H}(0)\dr^\dagger \vert 0\ra \nnb\\
&\equiv& -\ga g^{\mu\nu}-\frac{q^\mu q^\nu}{q^2}\dr\Pi^{(1)}_{\cal H}(q^2)+\frac{q^\mu q^\nu}{q^2} \Pi^{(0)}_{\cal H}(q^2)~,
 \label{eq:2-pseudo}
 \eea
where, e.g., in the case of the $D^*D$ configuration, the hadronic current reads\,:
\beq
{\cal O}^\nu_{\cal H}=(\bar c\gamma_\mu q)(\bar u\,i\gamma_5c)~.
\eeq
We have used the usual minimal duality ansatz:
  \bea
\hspace*{-0.65cm} \frac{1}{\pi}{\rm Im} \Pi_{\cal H}\hspace*{-0.1cm}&\simeq&\hspace*{-0.1cm} f_{\cal H}^2 M_{\cal H}^{8} \delta(t-M_{\cal H}^2) +\Theta(t-t_c) ``{\rm  Continuum}",
 \eea
 for parametrizing the molecule / four-quark state spectral function. $M_{\cal H}$ and $f_{\cal H}$ are the lowest ground state mass and coupling analogue to $f_\pi=131$ MeV. The ``Continuum" or ``QCD continuum"  is the imaginary part of the QCD correlator (as mentioned after Eq.\,\ref{eq:lsr}) from the  threshold $t_c$ which is assumed to smear all higher states contributions. This parametrization insures that both sides of the sum rules have the same large $t$ asymptotic behaviour which leads to the LSR in Eq.\,\ref{eq:lsr}. Within a such parametrization, one  obtains: 
 \beq
  {\cal R}^{c}_n\equiv {\cal R_H}\simeq M_{\cal H}^2~,
  \label{eq:mass}
  \eeq
 indicating that the ratio of moments appears to be a useful tool for extracting the mass of the hadron ground state\,\cite{SNB1,SNB2,SNB3}. The corresponding value of $t_c$  corresponds {\it approximately} to the mass of the 1st radial excitation. However, one should bear in mind that a such parametrization cannot distinguish two nearby resonances but instead will consider them as one ``effective resonance". 
 
\section{Optimization Criteria\label{sec:optim}}
 As $\tau$ (LSR variable), $t_c$ (QCD continuum threshold)  and $\mu$ (subtraction constant of the PT series) are free external parameters , we shall use stability criteria (minimum sensitivity on the variation of these parameters) to extract the hadron masses and couplings.
 \subsection{$\tau$-stability}
\begin{figure}[hbt]
\begin{center}
\hspace*{-7cm} {\bf a)} \\
\includegraphics[width=8cm]{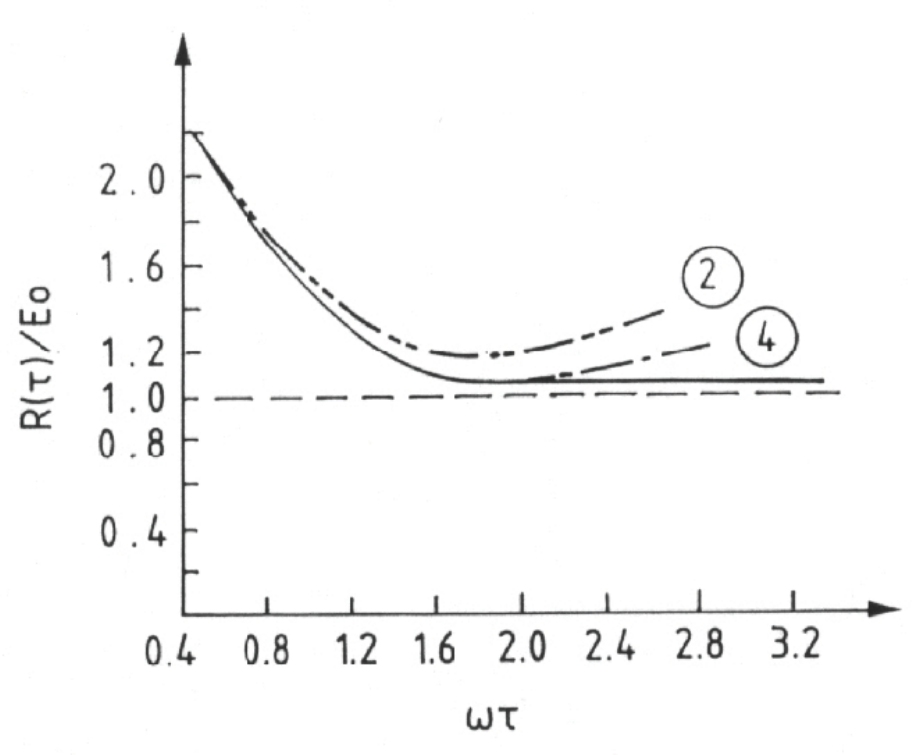} \\
\hspace*{-7cm}{ \bf b)} \\
\includegraphics[width=8cm]{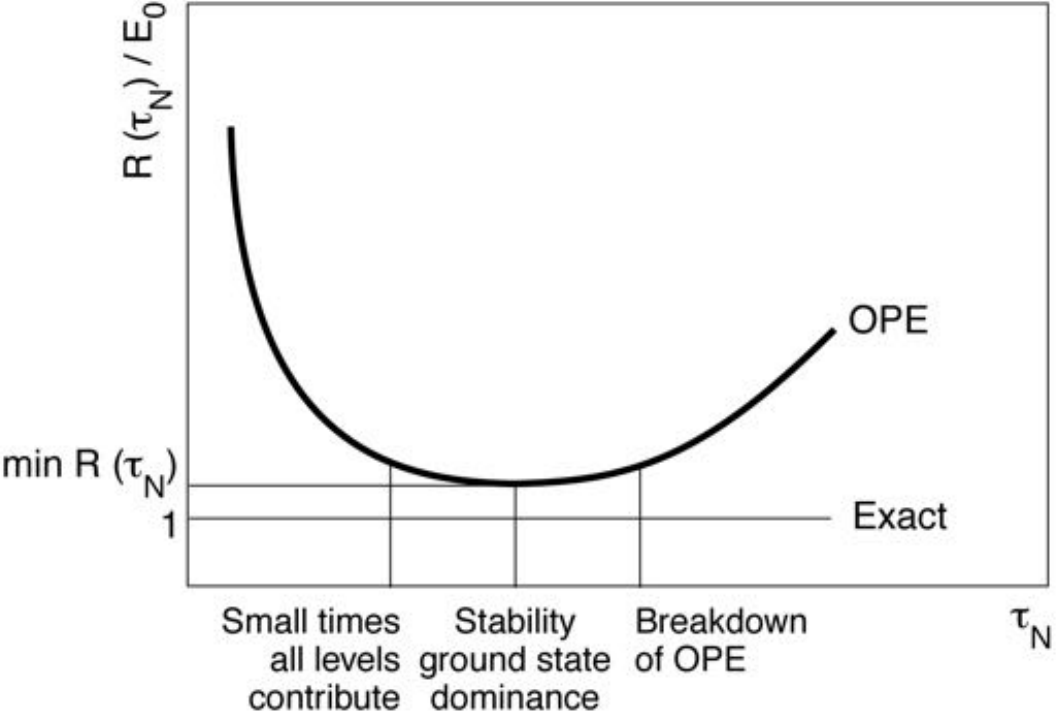}\\
{\hspace*{-7cm}\bf c)}\\
\hspace*{1cm}\includegraphics[width=9.5cm]{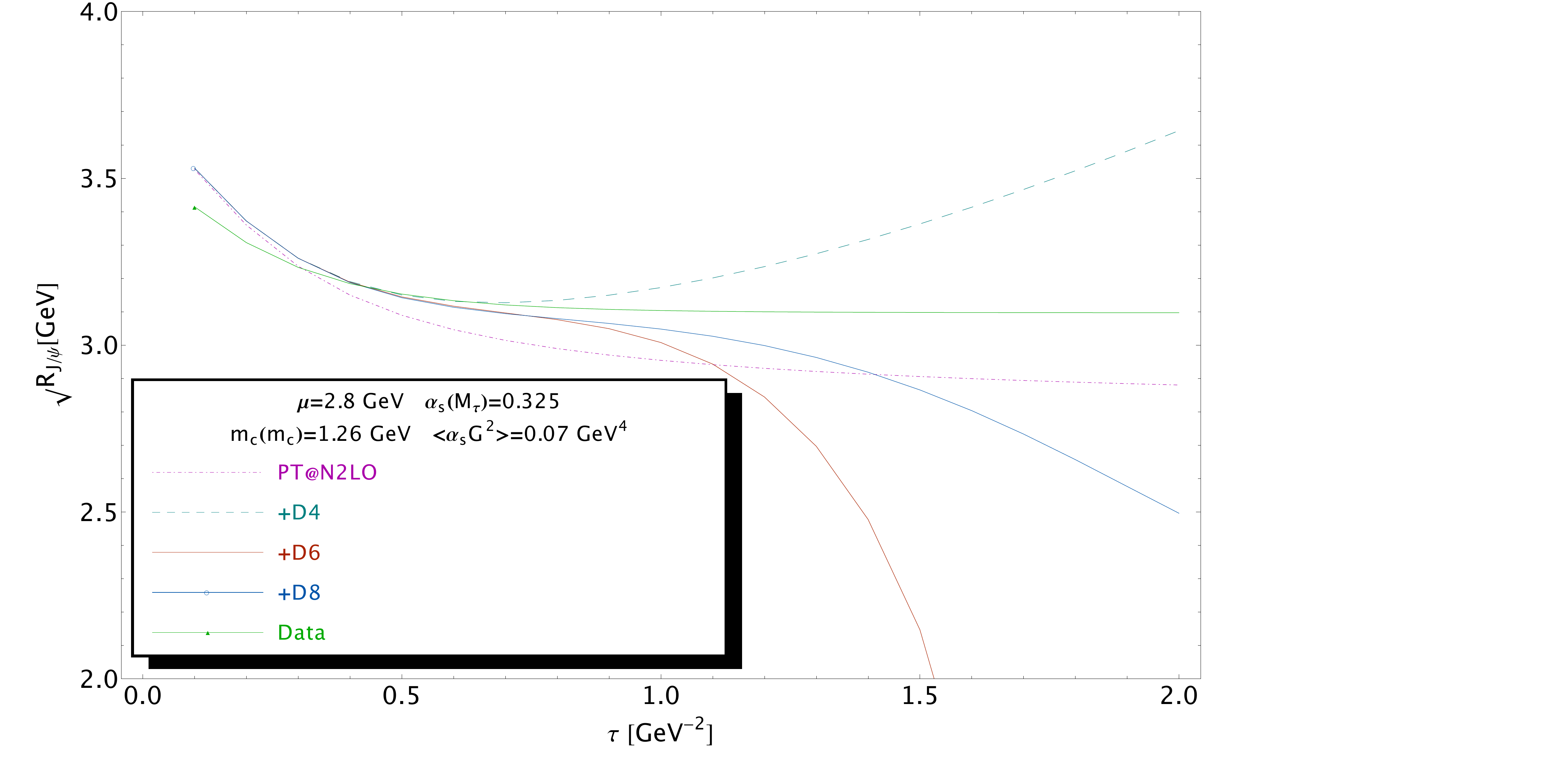}\\

\vspace*{-0.5cm}
\caption{\footnotesize  a) Harmonic oscillator state for each given truncation of the series compared to the exact solution (horizontal line); b) Schematic presentation of stability of the charmonium ratio of moments; c) Explicit analysis of the $J/\psi$ systems moment for different truncation of the OPE from e.g. \,\cite{SNparam,SNG}. } 
\label{fig:oscillo}
\end{center}
\vspace*{-0.5cm}
\end{figure}

 This optimization procedure for the case of the $\tau$-variable has been explicitly illustrated for the harmonic oscillator in quantum mechanics\,\cite{BELLa,BERTa} and from charmonium LSR analysis\,\cite{SNparam,SNG} (see Fig.\,\ref{fig:oscillo}) where the optimal result  is obtained at the minimum or inflexion point of the approximate series in $\tau$. These optimal values of $\tau$ are equivalent to the "so-called plateau" used in the literature using the Borel $M^2\equiv 1/\tau$ variable. However, one should note, e.g., in the case of $Z_c$, that the values of $M^2$ in Ref.\,\cite{WANG} move in a relatively small range $M^2\simeq (2.7\sim 3.3)$ GeV$^2\equiv \tau\simeq (0.30 \sim 0.37)$ GeV$^{-2}$ compared to the range of $\tau$-values analyzed in \,\cite{Zc}. 
\subsection{The $t_c$-stabiity}
\d The QCD continuum threshold $t_c$ is (in principle) a free parameter in the analysis though one (intuitively) expects it to be around the mass of the first excitation which cannot be accurate as the QCD continuum is supposed to smear all higher radial exctiations contributions to the spectral function. 

\d To be  conservative we take $t_c$ from the beginning of $\tau$-stability until the beginning  of $t_c$-stability\,\cite{SNB1,SNB2,SNB3} where the $t_c$-stability region corresponds to a complete dominance of the lowest ground state in the QSSR analysis. This conservative range of $t_c$-values is larger and wider than the usual choice done in the current literature where $t_c$ is taken at lower values of $t_c$ often below the beginning of the $\tau$-stability region. For the present case of $Z_c$, we obtain\,\cite{Zc}:
\beq
t_c=(22\sim 38)~{\rm GeV}^2,
 \label{eq:tc}
\eeq
 where the first value of $t_c$ corresponds to the beginning of $\tau$-minimum  of the coupling and the second one to the $t_c$-stability. 

\subsection{The $\mu$-stability}
 $\mu$-stability is used to fix in a rigorous optimal way, the arbitrary substraction constant  appearing in the PT calculation of the Wilson coefficients and in the QCD input renormalized parameters. We have obtained for $Z_c$\,\cite{Zc}:
\beq
\mu_c\simeq  4.65(5)~{\rm GeV}~,
\label{eq:mu}
\eeq
which has the same value as the one in our different analysis for the four-quark and molecule states\,\cite{MOLE16,SU3,Zb,DK,XTZ}. 

\d Alternatively, one can also eliminate the $\mu$-dependence of the result,  by working with the resummed quantity after applying the homogeneous Renormalization Group equation (RGE) obeyd the QCD expression of the LSR which is superconvergent\,:
\bea
\Big{\{}-\frac{\partial }{\partial t} +\beta(\alpha_s)\alpha_s \frac{\partial }{\partial \alpha_s}-\sum_i(1+\gamma_m(\alpha_s)\times\nnb\\
 x_i\frac{\partial }{\partial x_i}\Big{\}}{\cal L}^c_n(e^t \tau,\alpha_s,x_i,\mu)=0~,
\eea
where $t\equiv (1/2)L_\tau$, $x_i\equiv m_i/\mu$, $\beta$ is the $\beta$-function  and $\gamma_i$ is the quark mass anomalous dimension. The renormalization group improved (RGI) solution is:
\beq
{\cal L}^c_n(e^t \tau,\alpha_s,x_i)= {\cal L}^c_n(t=0,\bar\alpha_s(\tau), \bar x_i(\tau))~,
\eeq
where $\bar\alpha_s(\tau)$ and $\bar x_i(\tau)$ are the running QCD coupling and mass. 
However, the RGE solution $\mu^2=1/\tau$ would correspond to a lower value of $\mu\approx 1.6$ GeV where the convergence of the PT series can be questionable. An explicit comparison of the results from these two ways can be found in\,\cite{SNFB13}. 

Results based on these stability criteria have lead to successful predictions in the current literature (see\,\cite{SNB1,SNB2,SNB3}  and original papers). In the case of $Z_c$, we have obtained:
 \beq
 f_{Z_c}= 140(15)~{\rm keV}~,~~~~  M_{Z_c}= 3912(61)~{\rm MeV},
 \label{zcmass}
 \eeq
where $M_{Z_c}$ is in a remarkable agreement with the data $Z_c(3900)$ \,\cite{PDG}.


\section{The $Z_c$ ground state from FESR}
 In Ref.\,\cite{Zc} for a $D^*D$ molecule description of the $Z_c$, 
 the optimal values of the coupling and mass  have been extracted (see Eq.\,\ref{zcmass}) inside the conservative range of $t_c$ given in Eq.\,\ref{eq:tc}. The value of the mass does not present a  $\tau$-minimum but a  $\tau$-inflexion point where its value is about the same as the  one of the $\tau$-minimum of the coupling for the same value of $t_c$. 

To test the consistency of the values of the ground state coupling and mass extracted in this way from Laplace (global) sum rule (LSR), we use {\it local duality} Finite Energy Sum Rule (FESR). This approach has been extensively discussed in Ref.\cite{LAUNER} in the case of the $\rho$-meson where to NLO of PT series, the lowest moment gives the constraint :
\beq
\frac{M_\rho^2}{4\gamma_\rho^2}=\frac{t_c}{8\pi^2}\Big{[}1+\ga\frac{\alpha_s(t_c)}{\pi}\dr+{\cal O}(\alpha_s^2)\Big{]},
\label{eq:rho}
\eeq
for a minimal duality ansatz ``one resonance $\oplus$ QCD continuum
Using the experimental mass $M_\rho=775$ MeV and coupling $\gamma_\rho=2.55$, one obtains for $\alpha_s=0.39$:
\beq
\sqrt{t_c}\simeq 1.27~ {\rm GeV}
\label{eq:rhoprim}
\eeq
which is slightly lower than the mass 1465(25) MeV of the first radial excitation $\rho'$ of the $\rho$-meson. This value of $t_c$ is inside the stability region of the LSR analysis\,\cite{SNB1,SNB2}. 
  
\begin{figure}[hbt]
\vspace*{-0.25cm}
\begin{center}
\centerline {\hspace*{-7.5cm} \bf a) }
\includegraphics[width=7.2cm]{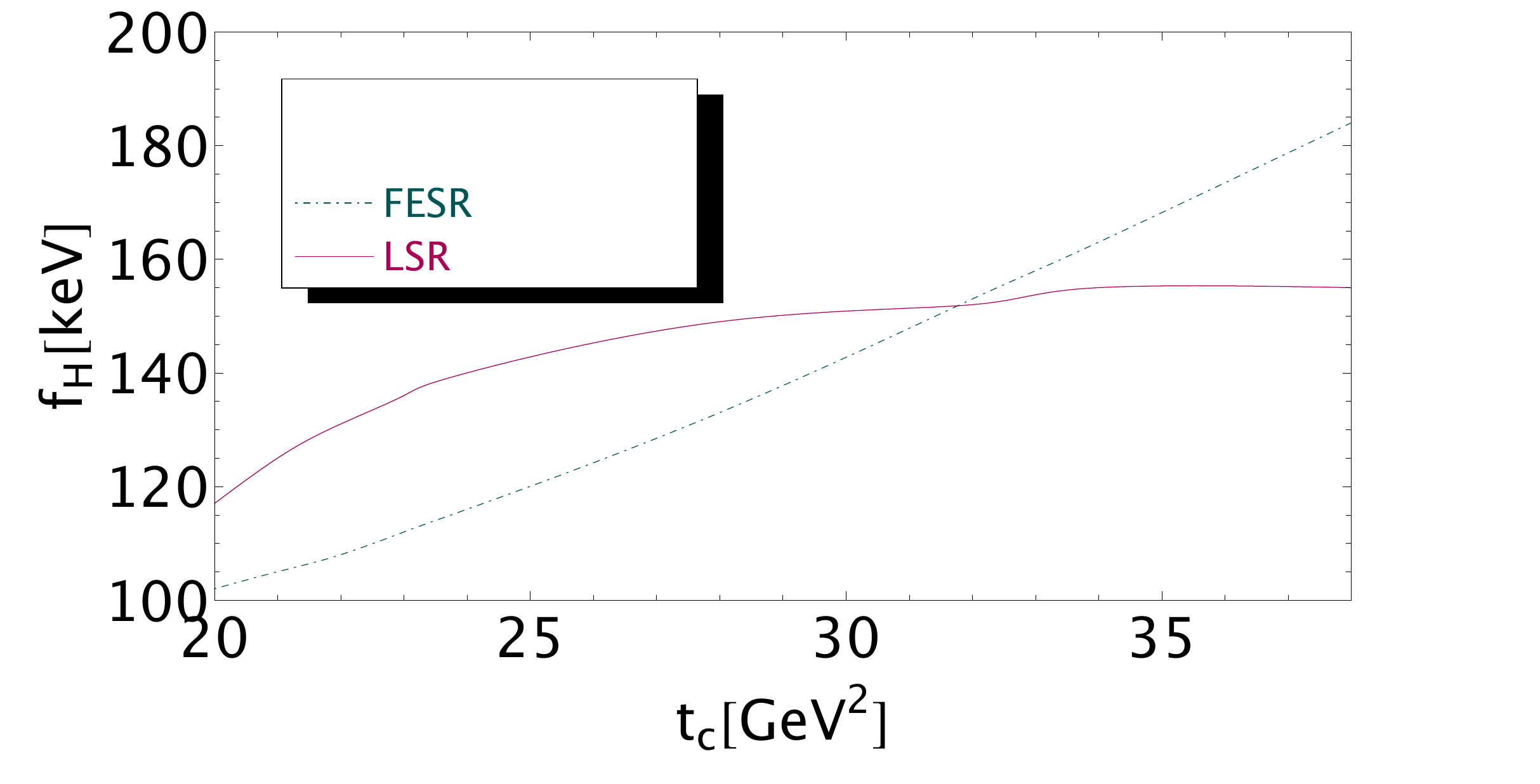}
\centerline {\hspace*{-7.5cm} \bf b) }
\includegraphics[width=7.2cm]{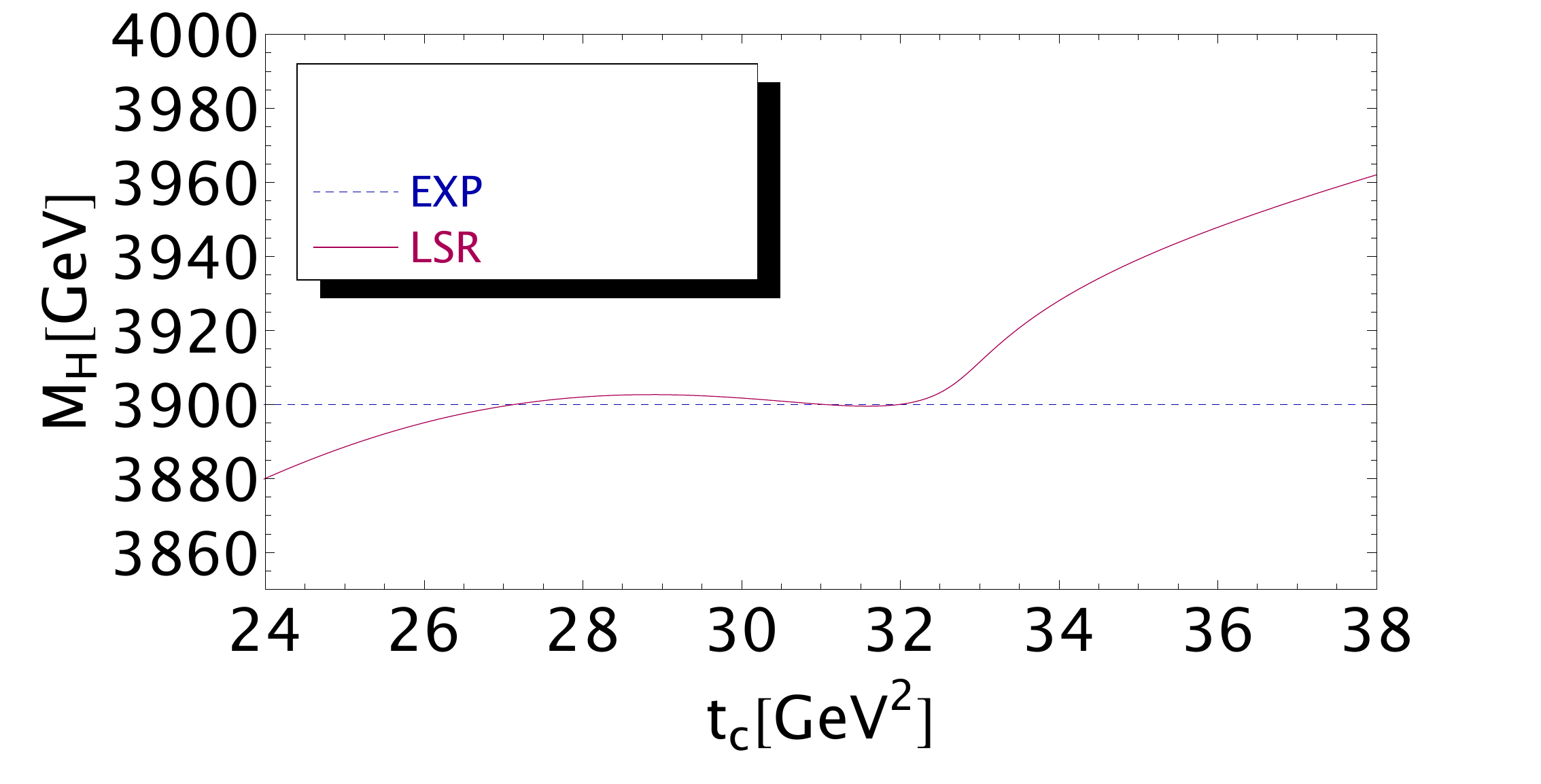}
\vspace*{-0.5cm}
\caption{\footnotesize  $Z_c$ parameters from LSR and FESR as a function of $t_c$ at NLO for $\mu$=4.65 GeV.} 
\label{fig:zc}
\end{center}
\vspace*{-0.25cm}
\end{figure}

 We extend this analysis to the case of the $Z_c$ meson assumed to be a $D^*D$ molecule. Using as input (in a first iteration) the mass the prediction: $M_{Z_c}=3912$ MeV in Table 3 of\,\cite{Zc}, we estimate $f_{Z_c}$. Then, we extract $M_{Z_c}$ using the ratio of moments at the $\tau$-minimum of $f_{Z_c}$ and at the corresponding value of $t_c$. In the 2nd iteration, we use this value of $M_{Z_c}$ to re-extract $f_{Z_c}$. We repeat this procedure for different $t_c$ for LSR.  The results are shown in Fig.\,\ref{fig:zc}
where a common stability region in $t_c$ is obtained for the coupling and the mass.  We notice that in the stability region, the experimental $Z_c$ mass is well reproduced from the LSR analysis. A similar procedure is done for FESR where, unlike LSR, the result increases with $t_c$. A similar behaviour has been obtained in the case of the $\rho$-meson (see the constraint in Eq.\,\ref{eq:rho}). The $t_c$-stability for FESR needs a complete data parametrization of the spectral function\,\cite{LAUNER} which is not yet possible for the $Z_c$.  One can also note that FESR overestimates the mass of $Z_c$ which is due to the fact that the 2nd moment entering in the ratio for extracting the mass is more affected by the higher mass radial excitations. The corresponding curve 
is not shown in Fig.\,\ref{fig:zc}. 

One can notice that the LSR and FESR predictions for the coupling meet at:
\beq
t_c\simeq 32~{\rm GeV}^2,
\label{eq:tcfesr}
\eeq
which is inside the conservative range in Eq.\,\ref{eq:tc} where\,:
 \beq
 f_{Z_c}\simeq 153(16)~{\rm keV}~,~~~~  M_{Z_c}\simeq  3900(60)~{\rm MeV},
 \label{fig:zcres}
 \eeq
 These values reproduce (within the errors) the ones in Ref.\,\cite{Zc}. 
\section{$Z_c$ radial excitations}
 If one attempts to identify the value of $t_c$ in Eq.\,\ref{eq:tcfesr} with the mass squared of the 1st radial excitation, one would obtain:
\beq
M_ {(D^*D)_1}\simeq 5657~{\rm MeV},
\label{eq:rad1}
\eeq
which we can identify with  $M_{(D^*D)_1}$=5709(70) extracted directly from LSR in Ref.\,\cite{Zc} with\,:
\beq
f_ {(D^*D)_1}=197(25)~{\rm keV}.
\label{eq:frad1}
\eeq

 In Ref.\,\cite{Zc}, we have also attempted to assume that the $Z_c(4430)$ is the 1st radial excitation of the $Z_c(3900)$. 
Then, we have estimated its coupling to the current to be\,:
\beq
f_ {(D^*D)_0}=46(56)~{\rm keV},
\label{eq:rad0}
\eeq
which is much smaller than the one of $(D^*D)_1$ in Eq.\,\ref{eq:frad1}. 

 We conclude from the previous study that the $(D^*D)_0$ and $(D^*D)_1$ states can be the radial excitations of the $Z_c(3900)$ having the parameters in Eqs.\,\ref{eq:rad1} to \ref{eq:rad0}. However, the $(D^*D)_0$ might have been masked by the $(D^*D)_1$ from the direct extraction using LSR due to its weaker coupling to the current. 

\section{On the four-quark condensates}
 Earlier estimates of the four-quark condensates:
\beq
\la 0\vert \bar \psi \Gamma_1 \psi \bar \psi \Gamma_2\psi\vert 0\ra
\eeq
($\Gamma_i$ is a generic notation for $\gamma$ matrices)
from $e^+e^-\to $ Hadrons data \,\cite{LNT,LAUNER}, $\tau$-decays\,\cite{SNTAU} and light baryon systems\,\cite{JAMI2a,JAMI2b,JAMI2c} have indicated a deviation of about a factor 3-4 of their value from vacuum saturation.  

 In Ref.\,\cite{WANG}, the author claims that, in the light meson systems, the effect of the four-quark condensate is relatively small compared to the lower dimension condensates one appearing in the OPE as it is multiplied by $\alpha_s$. This argument is not correct because, due to the anomalous dimension, the quantity $\alpha_s\la \bar \psi\psi\ra^2$ has a weak $\log^{1/9}(Q/\Lambda)$ behaviour for e.g. three light flavours (see e.g. \cite{SNB1,SNB2}). 
 
 The author in Ref.\,\cite{WANG} also claims that the corrections to the vacuum saturation is {\it obviously negligble} using an argument based on the eventual smallness of the $1/N_c$ corrections. In order to validate his claim, the author should compute explicitly the coefficient of such $1/N_c$ perturbative corrections and show that the non-perturbative contributions of hadronic intermediate states $\vert \pi\ra\la \pi\vert,~\vert \rho\ra\la \rho\vert\dots$ are negligible. He should also invalidate all previous phenomenological estimates of this quantity. 

\section{The OPE and PT series}
 In Ref.\,\cite{Zc}, the OPE is truncated at the dimension-six condensate contributions where the systematic error related to this truncation has been estimated by rescaling the dimension-six condensate contributions using the typical exponential factor $m_c^2\tau/3$ where the size of this estimate is about the one of the dimension-8 $\la\bar qq\ra \la \bar q Gq\ra$ condensate contributions obtained in\,\cite{MOLE16,SU3}. However, one should have in mind that this contribution is only a part of the complete $d$=8 condensate ones while the validity of the vacuum saturation used for its estimate is also questionable.  Therefore, a valuable claim on the convergence of the OPE requires an evaluation of the complete dimension-8 contributions and a non-use of factorization for estimating these high-dimension condensates which should mix under renormalization\,\cite{SNTARRACH}.  

 Alternatively, we use FESR to test the validity of the LSR results truncated at the dimension-six condensates\,\cite{Zc}.  Unlike the LSR where the OPE is done in terms of the $\tau$-variable, the OPE for FESR is done in terms of $t_c$ where its large value (see Eq.\,\ref{eq:tcfesr}) guarantees a much better convergence of the OPE which we illustrate for the coupling shown in Table\,\ref{tab:coupling}. As expected, we notice that the contributions of the high-dimension condensates are negligible while the one of the four-quark condensate is relatively large in this channel.  This result from FESR consolidates the one obtained from LSR in Ref.\,\cite{Zc} at lower scale.

   At the scale $\mu=4.65$ GeV, we also test the convergence of the PT series. For $t_c=32$ GeV$^2$, we obtain\,:
\beq
f_{Z_c}^{LO}=149.4~{\rm keV}~,~~~~f_{Z_c}^{NLO}=152.5~{\rm keV}~,
\label{eq:zc-lo}
\eeq
where the effect of the NLO correction is (almost) negligble. 
\begin{table}[hbt]
\setlength{\tabcolsep}{0.6pc}
\catcode`?=\active \def?{\kern\digitwidth}
 \begin{tabular}{cccccc}
 \hline
\hline
$Z_c$\,&$t_c$ [GeV]$^2$&$d_0$&$d_{0-4}$&$d_{0-5}$&$d_{0-6}$\\	\hline
$f_{Z_c}$ [keV]& 32&113.2&149.9&149.5&152.5\\
\hline\hline
\end{tabular}
{\scriptsize
 \caption{Perturbative (PT) ($d=0$) at NLO and non-perturbative condensate contributions of dimension $d\leq 6$ to the $Z_c$ coupling from local duality FESR. $d_{0-n}\equiv $ contributions of dimensions $d=0+(d\equiv n-1)+(d\equiv n)$ condensates. \label{tab:coupling}}
}
\end{table}

\section{On the analysis in Ref.\,\cite{WANG}}
The author in Ref.\,\cite{WANG} uses LSR within {\it his optimization procedure} to estimate the mass of ground state $Z_c(3900)$ and of the 1st radial excitation $Z_c(4430)$. He obtains:
\beq
M_{Z_c}=3.91^{+0.21}_{-0.17}~{\rm GeV},~~~~M_{Z'_c}=4.51^{+0.17}_{-0.09}~{\rm GeV}
\eeq
using the following favoured choice of parameters:
\subsection{Continuum threshold}
The author chooses the value $t_c=(22\sim 24)~\rm{GeV}^2$ for extracting the $Z_c(3900)$ and $Z_c(4430)$ masses and couplings. Hopefully, this value of $t_c$ is inside the  conservative stability region given in Eq.\,\ref{eq:tc}. 

\subsection{Plateau region and optimal results}
 The ``plateau region" is taken in the range $1/\tau\equiv T^2= (2.7\sim 3.3)~{\rm GeV}^2$ which is narrower [$\tau\sim (0.30\sim 0.37)$ GeV$^{-2}$] than the one in Ref.\,\cite{Zc} and in Fig.\,\ref{fig:pt}. One should remark that the scale of the figure in Ref.\,\cite{WANG} (and in some papers in the literature) is (exaggeratedly) enlarged which gives the impression of a large plateau.

 Taking the example of $f_{(D^*D)}$ in Fig.\,\ref{fig:pt}, one can remark that the minimum in $\tau$ obtained at LO becomes an inflexion point at NLO in the case of the vacuum saturation estimate of the four-quark condensate. It shows that the extraction of the optimal value does not necessary need 
a large plateau contrary to the claim in Ref.\,\cite{WANG}. The existence of a minimum or/and an inflexion point is sufficient for an approximate OPE and PT series according to the example of harmonic oscillator and charmonium channel discussed in Section\,\ref{sec:optim}.  
\subsection{Subtraction point $\mu$ and PT series convergence}
\begin{figure}[hbt]
\vspace*{-0.25cm}
\begin{center}
\centerline {\hspace*{-7.5cm} \bf a) }
\includegraphics[width=7.2cm]{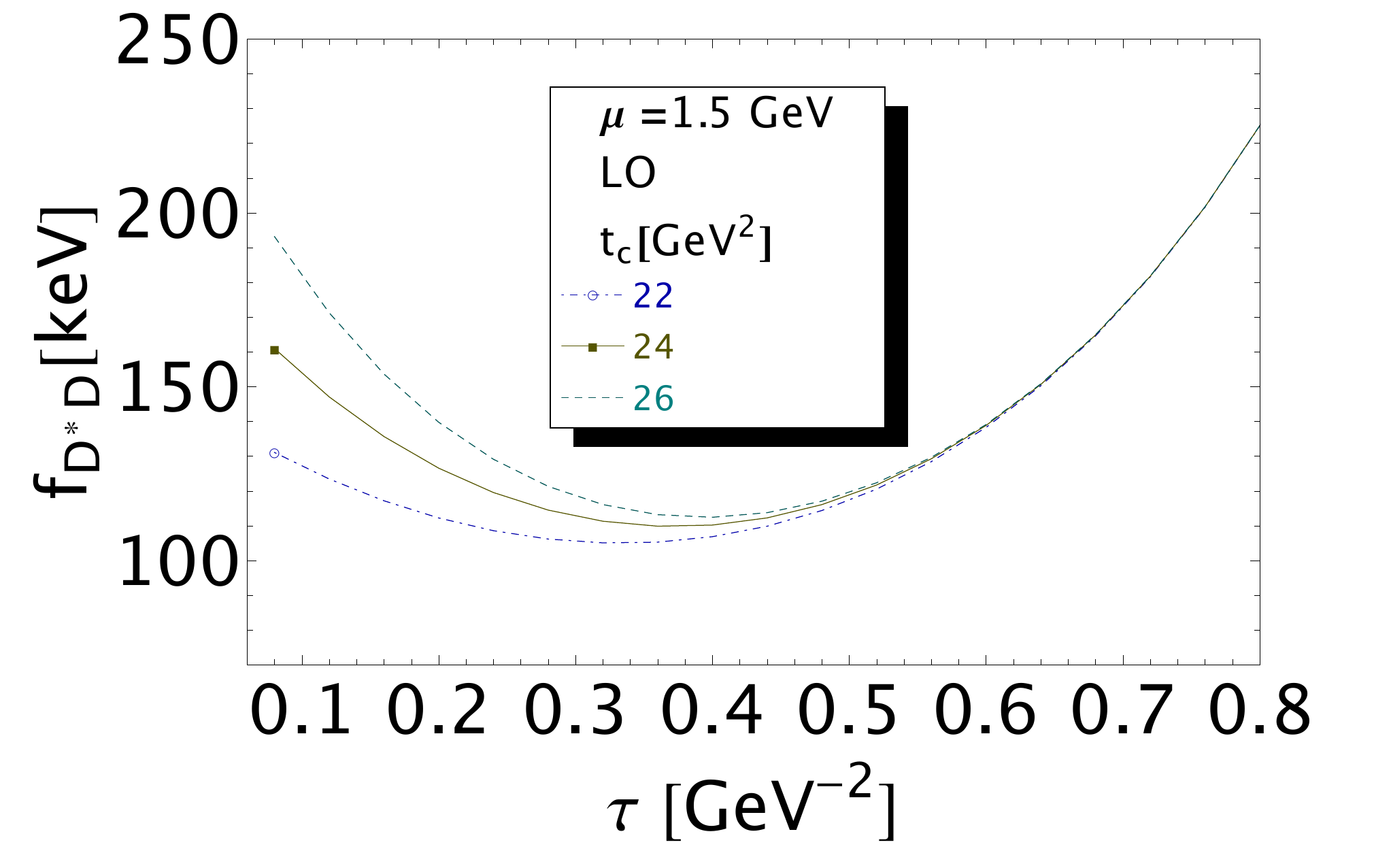}
\centerline {\hspace*{-7.5cm} \bf b) }
\includegraphics[width=7.2cm]{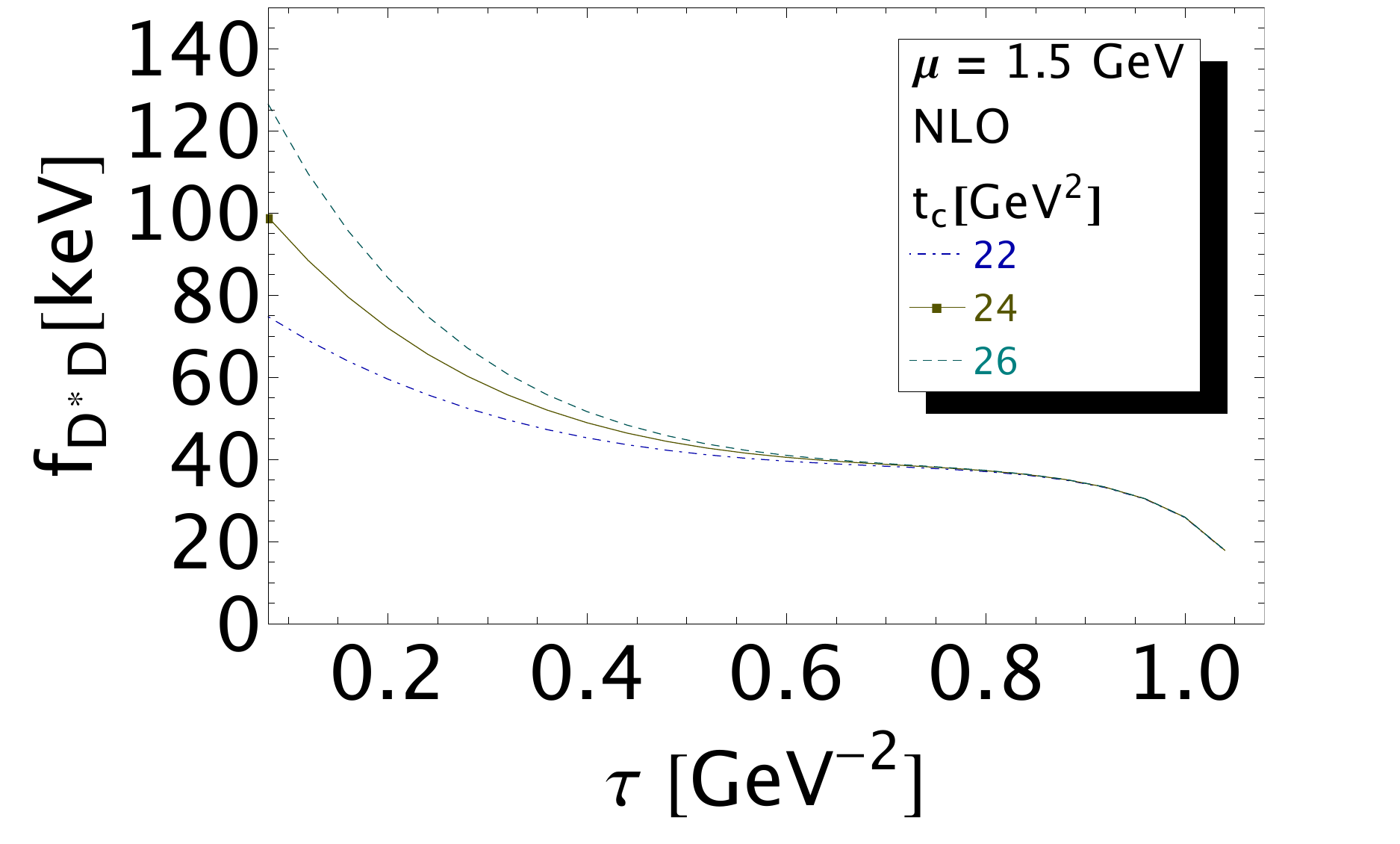}
\vspace*{-0.5cm}
\caption{\footnotesize  $f_{(D^*D)}$ as a function of $\tau$ at\,: a) LO and b) NLO for different values of $t_c$ and for $\mu$=1.5 GeV in the case of factorization of the four-quark condensate.} 
\label{fig:pt}
\end{center}
\vspace*{-0.25cm}
\end{figure}

The author favours the choice $\mu=1.5(2.7)$ GeV of the subtraction point  for extracting the $Z_c(Z'_c)$ masses and couplings. We check explicitly in Fig.\,\ref{fig:pt} the convergence of the PT series for extracting the $Z_c$ coupling and mass at $\mu=1.5$ GeV in the case of the factorization of the four-quark condensate used by the author in Ref.\,\cite{WANG}. 

From Fig.\,\ref{fig:pt} one can see that the NLO correction is huge for the favoured choice $\mu=1.5$ GeV of Ref.\,\cite{WANG}\,:
\beq
f_{D^*D}^{\rm LO}= 110(5)_{t_c}~{\rm keV},~~~~~~f_{D^*D}^{\rm NLO}= 41(1)_{t_c}~{\rm keV},
\eeq
for $t_c=24(2)$ GeV$^2$ and $\tau\simeq  0.4(0.6)$ GeV$^{-2}$ respectively for LO(NLO) where only the error induced by $t_c$ has been quoted. It indicates that the PT series is unreliable. 

\begin{figure}[H]
\vspace*{0.25cm}
\begin{center}
\centerline {\hspace*{-7.5cm} \bf a) }
\includegraphics[width=7.2cm]{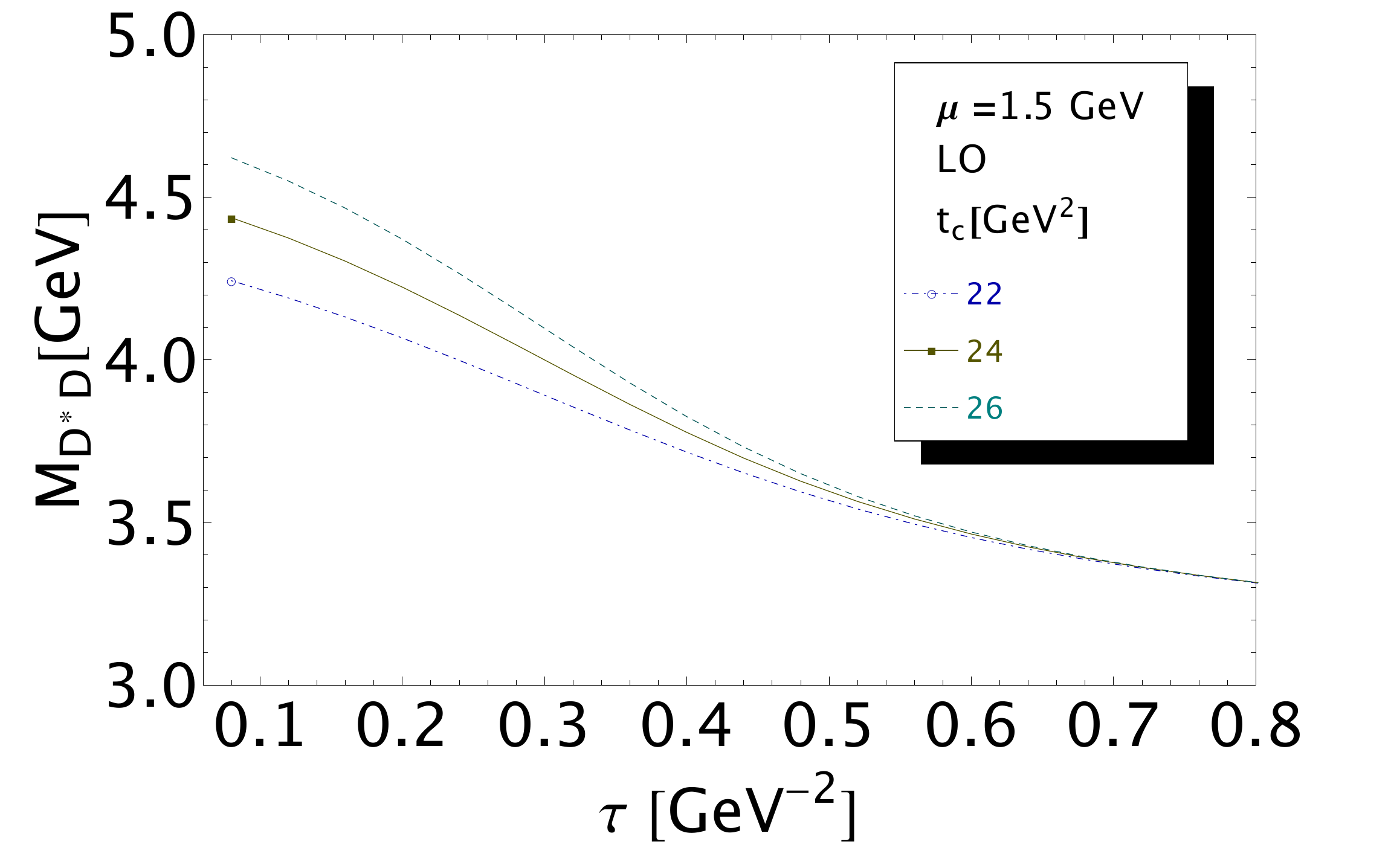}
\centerline {\hspace*{-7.5cm} \bf b) }
\includegraphics[width=7.2cm]{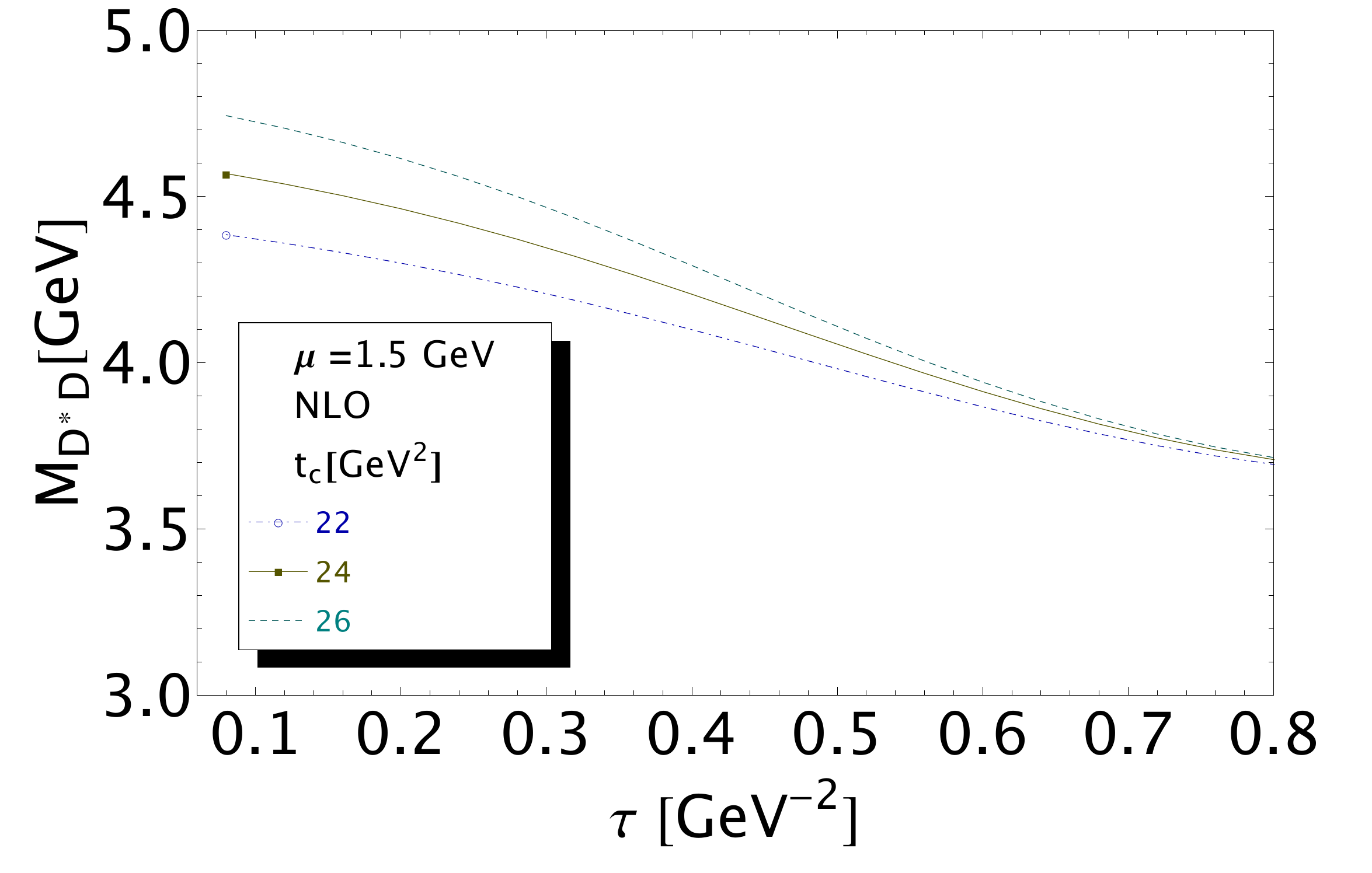}
\vspace*{-0.5cm}
\caption{\footnotesize  $M_{(D^*D)}$ as a function of $\tau$ at\,: a) LO and b) NLO for different values of $t_c$ and for $\mu$=1.5 GeV  in the case of factorization of the four-quark condensate.} 
\label{fig:mass}
\end{center}
\vspace*{-0.25cm}
\end{figure}

For the choice $\mu=2.7$ GeV used to extract the $Z_c(4430)$ parameter, the correction to the coupling of about 10\% is more reasonable.  In the case of the optimal value $\mu=4.65$ GeV obtained  in Ref.\,\cite{Zc} and in Eq.\,\ref{eq:zc-lo},  the correction to the coupling is (almost) negligible. The $\tau$-behaviour of the mass is shown in Fig.\,\ref{fig:mass} for $\mu=1.5$ GeV. One obtains in units of MeV\,:
\beq
M_{D^*D}^{ LO}= 3777(61)_{t_c},~~~~M_{D^*D}^{\rm NLO}= 3913(46)_{t_c}.
\eeq
The NLO corrections are moderate due to the cancellation of these contributions in the ratio of moments. However, this result obtained from unreliable individual expressions of the moments is misleading and should  not be (seriously) considered. 

\section{Summary and conlusions}
In this paper, we have used FESR to test the reliability of the LSR results\,\cite{Zc} within  the NLO corrections and where the OPE is truncated at the $d=6$ condensates. 

Compared to LSR, the OPE of FESR is more convergent while the $t_c$ (continuum threshold)-behaviour of the result does not present stability. The common solution of the two approaches shown in Fig.\,\ref{fig:zc} favours a value of $t_c$ around 32 GeV$^2$ which restricts the conservative $t_c$-range from LSR inside the $\tau$ to $t_c$ stability region given in Ref.\,\cite{Zc}. 

Attempting to identify this $t_c$-value with the mass of the  radial excitation $(D^*D)_1$, we obtain the one in Eq.\,\ref{eq:rad1} which confirms the direct LSR extraction in Ref.\,\cite{Zc}. 

Assuming that the $Z_c(4430)$ is the 1st radial excitation of the $Z_c(3900)$ (named $(D^*D)_0$ in Ref.\,\cite{Zc})  as expected from quark model\,\cite{QM} and from an extrapolation of $\psi'-J/\psi$ mass-splitting\,\cite{PDG}, we find the value 46(56) keV of its coupling\,\cite{Zc}. A such coupling is relatively weak compared to the one of the ground state 153 keV and of the 2nd radial excitation $(D^*D)_1$  of  197(25) keV extracted directly from LSR\,\cite{Zc}. This feature may explain why the $(D^*D)_0$ has been masked from a direct LSR analysis.  Using a Golberger-Treiman-like relation where the hadronic width behaves as $1/f_{\cal H}^2$, then, one may expect that the $Z_c(4430)$ is wider than the $Z_c(3900)$ as indicated by the data\,\cite{PDG}. 
 \input{bib_DstarD.tex}
\end{document}
